\documentclass[secnumarabic,aps,pra,amsfonts,twocolumn,%
showkeys,floatfix%
]{revtex4-1}

\usepackage{bm,amsfonts,amsthm}
\usepackage{natbib}
\usepackage{placeins}
\usepackage{longtable}
\usepackage{graphicx}
\newcommand{\gl}[1]{(\ref{#1})}

\begin{document}
\title{Constraining forces causing the Meissner effect}

\author{Ekkehard Kr\"uger} 

\affiliation{Institut f\"ur Materialwissenschaft,
  Materialphysik, Universit\"at Stuttgart, D-70569 Stuttgart, Germany}
%
\date{\today}
\begin{abstract}
  As shown in former papers, the nonadiabatic Heisenberg model
  presents a novel mechanism of Cooper pair formation which is not the
  result of an attractive electron-electron interaction but can be
  described in terms of quantum mechanical constraining forces. This
  mechanism operates in narrow, roughly half-filled superconducting
  bands of special symmetry and is evidently responsible for the
  formation of Cooper pairs in all superconductors. Here we consider
  this new mechanism within an outer magnetic field. We show that in
  the magnetic field the constraining forces produce Cooper pairs of
  non-vanishing total momentum with the consequence that an electric
  current flows within the superconductor. This current satisfies the
  London equations and, consequently, leads to the Meissner
  effect. This theoretical result is confirmed by the experimental
  observation that all superconductors, whether conventional or
  unconventional, exhibit the Meissner effect.
\end{abstract}

\maketitle

\section{Introduction}
The nonadiabatic Heisenberg model~\cite{enhm} (NHM) emphasizes the
picture of strongly correlated atomic-like electrons in nearly
half-filled narrow energy bands. Within the NHM, the appertaining
localized states are consequently represented by symmetry-adapted and
optimally localized Wannier functions. In some metals, these Wannier
functions must be chosen spin-dependent in order that they are both
symmetry-adapted and optimally localized~\cite{theoriewf}. An energy
band with such spin-dependent Wannier functions is called
``superconducting band'' because only metals possessing a narrow,
roughly half-filled superconducting band experimentally prove to be
(conventional, high-$T_c$ or other) superconductors, see the
Introduction of Ref.~\cite{theoriewf}. This observation can be
interpreted straightforwardly within the NHM~\cite{josn}. Within this
model, the formation of Cooper pairs is not the result of an
attractive electron-electron interaction but may be described in terms
of quantum mechanical constraining forces operating in superconducting
bands. There is evidence that these constraining forces are necessary
for the Hamiltonian of the system to possess superconducting {\em
  eigenstates}, see, e.g., Section 6 of Ref.~\cite{josybacuo7}. This
applies to all superconductors, whether conventional or
unconventional.

Also within the NHM, the formation of Cooper pairs is mediated by
bosons, which, however, bear the crystal spin angular momentum $1\cdot
\hbar$. More precisely, the electrons couple to the {\em energetically
  lowest} boson excitations of the crystal that possess the
crystal-spin angular momentum $1\cdot\hbar$ and are sufficiently
stable to transport it through the crystal~\cite{ehtc}. This
distinguishes the theory of superconductivity within the NHM from the
standard theory. The superconducting transition temperature $T_c$ is
determined by the excitation energy of the crystal-spin-1 bosons
mediating the pair formation. As is well-known, the kinetic energy of
particles is not changed by constraining forces (and, hence, they can
easily be overlooked). Thus, also in a superconducting band, $T_c$ is
determined by the standard theory of superconductivity. In particular,
in the isotropic elemental superconductors (often referred to as
``conventional'' superconductors) pure phonons are able to carry
crystal-spin-1 angular momentum~\cite{es,ehtc}. Thus, the transition
temperature of the elemental superconductors is still defined by the
Bardeen-Cooper-Schrieffer theory~\cite{bcs}.

The aim of this paper is to provide evidence that the constraining
forces causing the formation of Cooper pairs in superconducting bands
are also responsible for the Meissner effect.  When superconductors
are cooled below their transition temperature $T_c$, they not only
lose their electrical resistance but also create currents which
completely oppose an applied magnetic field. This second effect was
discovered 1933 by Meissner and Ochsenfeld~\cite{meissner} and is
generally referred to as Meissner-Ochsenfeld effect or, shortly,
Meissner effect. J.E. Hirsch~\cite{hirsch} argued that a mechanism
proposed to explain superconductivity must also explain the Meissner
effect because this effect is observed in all superconductors. We show
that the mechanism of Cooper pair formation defined within the NHM
meets this strict requirement of Hirsch.

However, we do not explain the Meissner effect but we
restrict ourselves to derive the London equations~\cite{london} which
are generally believed to explain the Meissner
effect~\cite{tinkham} (though they are partially called into question
by Hirsch, see Section~\ref{sec:conclusions}). In the following
Section~\ref{sec:cooperpaare} we briefly explain the mechanism of
Cooper pair formation within the NHM. In particular, we outline the
important role of both constraining forces and the time-inversion
symmetry in the formation of Cooper pairs.  In
Section~\ref{sec:innertimeinversion} we define the ``inner
time-inversion'' within an external magnetic field, in
Section~\ref{sec:totalmomentum} we will derive
Equation~\gl{eq:12} giving the total momentum of a Cooper pair in an
outer magnetic field, and in Section~\ref{sec:londonequations} we
shall derive the London equations.

\section{Cooper-pair formation in a superconducting band}
\label{sec:cooperpaare}
The mechanism of the Cooper pair formation in a narrow, roughly
half-filled superconducting band has been described in a former
paper~\cite{josn}. In this section we give a short overview of the
features of this mechanism necessary for an understanding of the
Meissner effect. For a more detailed summary see Section~3
of Ref.~\cite{josybacuo7}.

\subsection{Superconducting band in the absence of a magnetic field}
\label{sec:sband}
First we assume no outer magnetic field to be present.  The Bloch
functions of a superconducting band can be unitarily transformed into
optimally localized spin-dependent Wannier functions which are adapted
to the symmetry of the electron system~\cite{theoriewf}. In this
context, the ``symmetry of the electron system'' also comprises the
time-inversion symmetry.  The NHM defines atomic-like electrons with
localized states represented by these spin-dependent Wannier
functions.  As a consequence of their spin dependence, the spin
directions of the Bloch states are {\bf k} dependent in the ground
state of a narrow, roughly half-filled superconducting band (this
striking feature of the Bloch electrons suggests interpreting
superconductivity as ``{\bf k} space magnetism''~\cite{ekm}).  The
Bloch functions $\varphi_{{\bf k}, q, {\bf m}}({\bf r}, t)$ are
labeled, as usual, by the wave vector {\bf k} and the band index $q$,
but no longer by the electron spin {\bf s} since the spin direction is
{\bf k} dependent. They are rather labeled by the ``crystal spin''
{\bf m} defined within the NHM~\cite{josn,theoriewf}.

In a system with {\bf k} dependent spin directions the electrons
couple to crystal-spin-1 boson excitations in order that the total
crystal-spin angular-momentum is conserved during the ever-present
scattering processes in the electron system, see Section~3.2 of
Ref.~\cite{josybacuo7}. 
At low temperatures, the electrons try to occupy a state in which the
electrons alone satisfy the conservation of spin-angular momentum.
The only fixed spin directions in a superconducting band are those of
a Bloch state $\varphi_{{\bf k}, q, {\bf m}}({\bf r}, t)$ and its
time-inverted state,
\begin{equation}
  \label{eq:24}
\varphi_{-{\bf k}, q,
    -{\bf m}}({\bf r}, t) = K\varphi_{{\bf k}, q,
    {\bf m}}({\bf r}, t),
\end{equation}
since both states have exactly opposite spin directions. $K$ denotes
the operator of time inversion.

At low temperatures, the electrons form Cooper pairs consisting in
each case of a Bloch state and its time inverted state. When all the
electrons of the superconducting band form Cooper pairs with zero
total spin-angular momentum, the conservation of spin angular-momentum
is satisfied in the electron system alone, see the group-theoretical
substantiation in Section 3.2 of Ref.~\cite{josybacuo7}.

The mechanism of Cooper pair formation can be described in terms of
constraining forces produced by the crystal-spin-1 boson excitations,
see Section 3.3 of Ref.~\cite{josybacuo7}. As illustrated in Fig.~3 of
Ref.~\cite{josi}, these constraining forces behave like classical
constraining forces produced by springs: Let ${\cal P}$ be the Hilbert
space spanned by the electron states in the superconducting band and
${\cal P}^0$ the subspace of ${\cal P}$ in which all the electrons
form Cooper pairs. Assume all the electrons of the superconducting
band initially to be in ${\cal P}^0$. Whenever two electrons are
scattered out of ${\cal P}^0$, a crystal-spin-1 boson pair is excited
which can only be reabsorbed when the electrons are scattered in such
a way that again they lie in ${\cal P}^0$.  Hence, the crystal-spin-1
bosons behave like ``springs'' that push the electrons back into
${\cal P}^0$. So we may speak of ``spring-mounted'' Cooper pairs.

\subsection{Superconducting band in an outer magnetic field}
\label{sec:sbandmagneticfield}
Now assume an outer magnetic field to be switched on.  An absolutely
consistent mathematical description of superconductivity in an outer
magnetic field would require to show that the spin-dependent Wannier
functions in a superconducting band may be chosen symmetry-adapted
even in the presence of an outer magnetic field, as it has been
carefully established~\cite{theoriewf} for the field-free case.
Though the symmetry of the Bloch and Wannier functions is, in
principle, known in magnetic fields~\cite{brown,wannier,fischbeck},
this would be a complicated and, as I believe, physically needless
task. Instead, we should keep in mind that the spin-dependent Wannier
functions represent localized electron states that {\em really 
  exist} in the material. These localized states clearly are adapted
the symmetry of the electron system. For this reason we can assume
that the spin-dependent Wannier functions in a superconducting band
may be chosen adapted to the symmetry of the electron system even in
the presence of an outer magnetic field.  In this context, the
symmetry of the electron system comprises the inner time-inversion
symmetry as shall be defined in Section~\ref{sec:totalmomentum}.

\section{The Hamiltonian in an uniform magnetic field}
The Hamiltonian of an electron in a solid state and in a uniform
external magnetic field has the form
\begin{equation}
  \label{eq:1}
\mathcal{H} = \frac{1}{2m}\big({\bf p} + \frac{e}{c}{\bf A}\big)^2 +
V({\bf r}),
\end{equation}
where
\begin{equation}
  \label{eq:4}
  {\bf p } = {\bf p}_{kin}-\frac{e}{c}{\bf A}
\end{equation}
is the operator of the generalized momentum, $m$ is the electron mass,
$e$ the proton charge, $V({\bf r})$ is the periodic potential,
${\bf p}_{kin}$ is the so-called ``kinetic momentum'', and ${\bf A}$
denotes the operator of the vector
potential~\cite{kittel,landaud}. An additional term standing for the
energy of the electron spins in the magnetic field is neglected.

The translation operators in the magnetic field may be written as
\begin{equation}
  \label{eq:18}
  T({\bf R}) = e^{-i{\bf R}\cdot {\bf p}/\hbar},
\end{equation}
where {\bf R} is a lattice vector and {\bf p} is the generalized
momentum given in Equation~\gl{eq:4}~\cite{brown}. Since the
translation operators $T({\bf R})$ commute with $\mathcal{H}$~\cite{brown},
\begin{equation}
  \label{eq:20}
  [T({\bf R}), \mathcal{H}] = 0, 
\end{equation}
we may label the eigenfunctions
of $\mathcal{H}$ by the generalized impulse {\bf p } and write
\begin{equation}
  \label{eq:2}
  \mathcal{H}\varphi_{{\bf p}, q, {\bf m}}({\bf r}, t) = E_{{\bf p}, q, {\bf m}}\varphi_{{\bf p}, q, {\bf m}}({\bf r}, t),
\end{equation}
as it was already performed by Onsager to interpret the de Haas-van
Alphen Effect~\cite{onsager}. $q$ still is the band index and $t$ is
the spin coordinate.  Just as in the field-free case, {\bf m} does not
stand for the electron spin but denotes the crystal spin since the
spin direction depends on {\bf p} in a narrow, roughly half-filled
superconducting band.

\section{Cooper pairs within an outer magnetic field}

\subsection{The inner time-inversion} 
\label{sec:innertimeinversion}
Consider a superconducting sample within an external magnetic field
generated by Helmholtz coils fare away from the sample.  As is
well-known, the electron system within the sample is invariant under
time inversion only if additionally the magnetic field ${\bf B}$ and,
hence, the vector potential ${\bf A}$ is inverted,
\begin{equation}
  \label{eq:16}
  K^{-1}{\bf A}K = -{\bf A},
\end{equation}
where $K$ denotes the operator of time inversion, see, e.g.,
Ref.~\cite{landaud}. This important phenomenon can be understood
already in classical physics: in a magnetic field, the Lorentz force
generates within the sample a circular motion of the electrons. An
inversion of the time of the system produces a circular motion of the
opposite direction of rotation. In a fixed magnetic field, however,
the Lorentz force generates in any case circular motions of the same
sense of rotation. Hence, a time inverted circular motion of the
electrons may exist only in the inverted magnetic field. Thus, an
inversion of the time requires that the experimentalist additionally
reverses the polarity of the battery connected with the Helmholtz
coils. Hence, $K$ is not a symmetry operation of the electron system.

This problem has been overcome for special sheared
solids~\cite{bonella} and for reversible microscopic
systems~\cite{dalcengio}. In the present paper, however, we do not
consider the standard time-inversion represented by $K$ connected with
the complete system consisting of both the superconducting sample and
the Helmholtz coils. Instead, we see the superconducting sample as an
inner isolated system within a fixed magnetic potential {\bf A}
produced by the outer Helmholtz coils. We define an operator
$\overline{K}$ inverting the time $\tau$ within the inner electron
system,
\begin{equation}
  \label{eq:14}
  \tau \rightarrow -\tau,
\end{equation}
without changing the outer magnetic field, 
\begin{equation}
  \label{eq:26}
  \overline{K}^{-1}{\bf A}\overline{K} = {\bf A}.  
\end{equation}
Thus, this operator $\overline{K}$ of the ``inner time inversion'' has
the same effect as $K$~\cite{wigner} on the kinetic momenta ${\bf
  p}_{kin}$, the spins {\bf s} and the positions ${\bf r}$ of the
inner electrons,
\begin{eqnarray}
  \overline{K}^{-1}{\bf p}_{kin}\overline{K} &=& -{\bf p}_{kin},\label{eq:8}\\
  \overline{K}^{-1}{\bf s}\overline{K} &=& -{\bf s},\label{eq:11}\\ 
  \overline{K}^{-1}{\bf r}\overline{K} &=& {\bf r}.\label{eq:13}
\end{eqnarray}
In contrast to the standard time inversion $K$, however, it does not
invert the sense of rotation of the circular motions produced by the
outer Lorentz force. Also $\overline{K}$ is an anti-linear operator
because it complies with the conditions given in Section 26 in the
textbook of E. P.~Wigner~\cite{wigner}.

\subsection{The total momentum of a Cooper pair}
\label{sec:totalmomentum}
With Equation~\gl{eq:4} the Hamiltonian may be written as  
\begin{equation}
  \label{eq:9}
  \mathcal{H} = \frac{1}{2m}\big({\bf p}_{kin})^2 +
  V({\bf r})
\end{equation}
showing immediately that $\overline{K}$ commutes with $\mathcal{H}$,
\begin{equation}
  \label{eq:7}
  \overline{K}^{-1}\mathcal{H}\overline{K} = \mathcal{H},
\end{equation}
if we continue to neglect the energy of the electron spins in the
magnetic field. From this result follows the significant insight that
the inner time-inversion $\overline{K}$ {\em is a symmetry operation
  of the inner electron system}. 

As argued in Section~\ref{sec:sbandmagneticfield}, the magnetic
Wannier functions are adapted to the inner time-inversion just as they
are adapted to the standard time inversion in the field-free case. As
a consequence, the operator $\overline{K}$ acts on the crystal spin
{\bf m} in the same way as it acts on the spin {\bf s},
\begin{equation}
  \label{eq:25}
  \overline{K}^{-1}{\bf m}\overline{K} = -{\bf m}, 
\end{equation}
as it has been shown for the zero-field case in Section 7.3.1 of
Ref.~\cite{theoriewf}.

Since the operator $\overline{K}$ commutes with $\mathcal{H}$,
$\overline{K}$ transforms an eigenstate $\varphi_{{\bf p}, q, {\bf
    m}}({\bf r}, t)$ of $\mathcal{H}$ into a new eigenstate of
$\mathcal{H}$,
\begin{equation}
  \label{eq:5}
  \varphi_{{\bf p}', q,-{\bf m}}({\bf r}, t)  =
  \overline{K}\varphi_{{\bf p}, q, {\bf m}}({\bf r}, t),  
\end{equation}
associated with the same energy,
where
\begin{equation}
  \label{eq:6}
{\bf p}' = \overline{K}^{-1}{\bf p}\overline{K}. 
\end{equation}
With Equations~\gl{eq:4},~\gl{eq:26} and~\gl{eq:8} we obtain
\begin{equation}
  \label{eq:10}
  {\bf p}' = -{\bf p}_{kin}-\frac{e}{c}{\bf A}.
\end{equation}

Remember that the direction of the electron spins depends on {\bf p} in
a narrow, roughly half-filled superconducting band. Just as in the
field-free case, the constraining forces produced by the
crystal-spin-1 excitations generate Cooper pairs with {\em exactly}
vanishing total spin-angular momentum. Equation~\gl{eq:11} ensures
that the spins of the two electrons occupying the states
$\varphi_{{\bf p}', q, -{\bf m}}({\bf r}, t)$ and $\varphi_{{\bf p},
  q, {\bf m}}({\bf r}, t)$ in Equation~\gl{eq:5} are exactly
anti-parallel. Consequently, these two states (and only these two
states) can form  Cooper pairs. (The basic Equation~(125) of
Ref.~\cite{theoriewf} ensuring a vanishing total spin-angular momentum
is satisfied even in an outer magnetic field if we replace ${\bf k}$
by ${\bf p}$, ${-\bf k}$ by ${\bf p'}$, and $K$ by $\overline{K}$ in
the derivation of this equation.)

 Hence, in a magnetic field, the total momentum ${\bf
  p}_c$ of the two electrons forming a Cooper pair in a
superconducting band does not vanish, but has the value
\begin{eqnarray}
  \label{eq:12}
  {\bf p}_c &=& {\bf p} + {\bf p}'\nonumber\\
  &=& -\frac{2e}{c}{\bf A}.
\end{eqnarray}
This equation gives the exact total momentum of a Cooper pair within
an outer magnetic field. It shall be interpreted in the following
Section~\ref{sec:londonequations}.

\section {The London Equations}
\label{sec:londonequations}
Equation~\gl{eq:12} shows that the kinetic momenta of the two Bloch
states forming a Cooper pair cancel each other.  However, the term
$-\frac{2e}{c}{\bf A}$ indicates that the Lorentz force still is
active and forces the two electrons to perform a circular motion with
the same sense of rotation each. Because the two electrons move on
different orbitals, the probability to meet an electron at a certain
position {\bf r} is different for the two electrons and, hence, their
average total kinetic momentum \text{$<\!{\bf p}_{c, kin}\!>$} at {\bf
  r} needs not vanish.  Thus, the electron pair with the momentum ${\bf
  p}_c$ may produce a ${\bf r}$ dependent electrical current ${\bf
  j_c}$ which is defined by the symmetry of the system.

To determine ${\bf j_c}$, we rewrite Equation~\gl{eq:12} as
\begin{equation}
  \label{eq:17}
  {\bf p}_c = -\frac{e}{c}{\bf A} -\frac{e}{c}{\bf A}
\end{equation}
showing that ${\bf p}_c$ has the form given in Equation~\gl{eq:4} if we
interpret one of the addends as the average kinetic momentum
\begin{equation}
  \label{eq:22}
  <\!{\bf p}_{c, kin}\!>\  =  -\frac{e}{c}{\bf A}
\end{equation}
of an one-electron state.

Due to this interpretation~\gl{eq:22}, the operator
\begin{equation}
  \label{eq:21}
   T({\bf R}) = e^{-i{\bf R}\cdot {\bf p}_c/\hbar}
\end{equation}
becomes a translation operator commuting with $\mathcal{H}$, and,
hence, the one-electron state with the momentum ${\bf p}_c$ becomes an
eigenstates of $\mathcal{H}$.  Consequently, an electrical current
represented by this state has physical reality.

Thus, the contribution of one Cooper pair to the electric current
amounts to
\begin{eqnarray}
  \label{eq:15}
  {\bf j_c}\ &=&\frac{e}{m}<\!{\bf p}_{c, kin}\!>\ \nonumber\\
         &=&-\frac{e^2}{mc}{\bf A}.
\end{eqnarray}
${\bf j_c}$ is invariant under the inner time-inversion $\overline{K}$
because it is originally defined by Equation~\gl{eq:12}, i.e., by the
outer vector potential {\bf A}.

Equation~\gl{eq:15} is the result of this paper. It contains both
London equations~\cite{london} in a compact form, see Equation (1.8)
in the textbook of M. Tinkham~\cite{tinkham}.

\section{Conclusions}
\label{sec:conclusions}
This paper provides evidence that the constraining forces causing the
formation of Cooper pairs in narrow, roughly half-filled
superconducting bands are also responsible for the Meissner effect. In
the framework of the nonadiabatic Heisenberg model, the Meissner effect
is an intrinsic part of superconductivity.

Hirsch~\cite{hirsch} argues that neither BCS theory nor London
electrodynamic theory describes superconductivity. But he adds that
parts of both BCS theory and London theory are undoubtedly
correct. From my point of view, I can confirm this strong statement of
Hirsch. However, I specify that BCS theory as well as London theory
are correct if the constraining forces operating in narrow, roughly
half-filled superconducting bands are present.
 
\acknowledgements I am very indebted to Guido Schmitz for his support
of my work.


\end{document}